\documentclass[prb,twocolumn,draft,amsmath,showpacs]{revtex4}
\usepackage{graphics}

\begin{document}

\bibliographystyle{prsty}
\input epsf

\title {Infrared optical properties of the spin-1/2 quantum magnet
$TiOCl$}

\author {G. Caimi}
\affiliation{Laboratorium f\"ur Festk\"orperphysik, ETH
Z\"urich,
CH-8093 Z\"urich, Switzerland}\

\author {L. Degiorgi}
\affiliation{Paul Scherrer Institute, CH-5232 Villigen and
Laboratorium f\"ur Festk\"orperphysik, ETH Z\"urich,
CH-8093 Z\"urich, Switzerland}\

\author {N.N. Kovaleva and P. Lemmens}
\affiliation{Max Planck Institute for Solid State Research, Heisenbergstr.
1, D-70569 Stuttgart, Germany}\

\author {F.C. Chou}
\affiliation{Center for Materials Science and Engineering, M.I.T.,
Cambridge, MA 02139, U.S.A.}\

\date{\today}

\begin{abstract}
We report results on the electrodynamic response of $TiOCl$, a
low-dimensional spin-1/2 quantum magnet that shows a spin gap formation
for T$<T_{c1}$= 67 $K$. The Fano-like shape of a few selected infrared
active phonons suggests an interaction between lattice vibrations and a
continuum of low frequency (spin) excitations. The temperature
dependence of the phonon mode parameters extends over a broad
temperature range well above $T_{c1}$, indicating the presence of an
extended fluctuation regime. In the temperature interval between 200 $K$
and $T_{c1}$ there is a progressive dimensionality crossover (from two
to one), as well as a spectral weight shift from low towards high
frequencies. This allows us to identify a characteristic energy scale of
about 430 $K$, ascribed to a pseudo spin-gap.
\end{abstract}

\pacs{78.20.-e, 71.36.+c}
\maketitle

\section{Introduction}
Low-dimensional quantum spin systems, based on complex transition metal
oxides, recently attracted a lot of attention, particularly as a fascinating
playground to study spin-charge separation, spin-gap states and quantum
disorder. Proposals, that the exotic properties of low-dimensional spin-1/2
quantum magnets might also play a major role in shaping the mechanism for high
temperature superconductivity, led to a vigorous experimental
activity on materials involving $Cu^{2+}$ ions with a 3d$^{9}$ configuration
(S=1/2). Other examples of S=1/2 are notably $Ti^{3+}$ and $V^{4+}$ systems
in d$^{1}$ configuration (i.e., one single d-electron occupies one of the
t$_{2g}$ orbitals). In this respect, the layered $TiOX$ ($X=Cl$ and $Br$)
compounds are most promising and are candidates for exotic electronic
configurations as in the resonating-valence-bond (RVB) model \cite{beynon} and for superconductivity
based on dimer fluctuations \cite{seidel2,seidel}.

$TiOCl$ has been first considered as a two-dimensional antiferromagnet, an
electron analog to the high-temperature cuprates. This was based on 
considerations about the electronic properties and on the bi-layer structure of the compound formed
by edge-sharing, distorted $TiO_4Cl_2$ octahedra \cite{seidel}. High quality single
crystals of $TiOCl$ display a kink in the spin susceptibility $\chi(T)$
below about $T_{c2}$=94 $K$, followed by a pronounced drop at $T_{c1}$=67
$K$. This leads to a non-magnetic ground state at $T_{c1}$, therefore
signaling the opening of a singlet-triplet spin-gap \cite{seidel,imai}. It
has been proposed that the effective dimensionality of the $TiO$ layers is
reduced from two to one by an orbital ordering at $Ti^{3+}$ sites and that
the singlet ground state is reached by a phase transition also involving
lattice degrees of freedom. Consequently, the spin gap formation and the
associated fluctuations in $TiOCl$ emerge in an entirely new perspective
from a quantum antiferromagnet and a scenario based on a spin-Peierls (SP)
transition occurring below $T_{c1}$ has been accredited as the most
plausible interpretation of the experimental findings \cite{seidel}.

The title compound seems to be also an ideal material to investigate a
broken symmetry ground state with orbital degrees of freedom but without
charge ordering. The relevant role played by large electronic energy scales,
associated to the orbital degrees of freedom, differentiates $TiOCl$ from
$CuGeO_{3}$ and $NaV_{2}O_{5}$, two other intensively studied spin-Peierls
systems: $CuGeO_{3}$ being characterized by a state without orbital and
charge degrees of freedom, $NaV_{2}O_{5}$ with orbital and charge degrees of
freedom \cite{imai,smolinski, mostovoy,lemmensrev}. In this context, it is worth mentioning that the
coupling of orbital to spin degrees of freedom in a chain system may
establish a novel route to the formation of spin gap states and spin-orbital
excitations \cite{pati,yamashita,koleshuk}.

Optical methods, like infrared or Raman spectroscopy, are powerful
experimental tools in revealing the characteristic energy scales associated
to the development of broken symmetry ground states, driven by magnetic
and/or structural phase transitions. Indeed, information on the nature of
the electronic (magnetic) ground state, lattice distortion and interplay of
electronic (magnetic) and lattice degrees of freedom can be obtained
studying in detail the electronic (magnetic) excitations and the phonon
spectrum, as a function of temperature. Furthermore, the role played by
strong quantum fluctuations in S=1/2 systems, an issue of key importance,
can be addressed as well. We provide here a complete set of infrared optical
data on $TiOCl$ and a thorough analysis of its electrodynamic response. The
paper is organized as follows: we first briefly describe the experiment and
present the data; the discussion will then emphasize the temperature
dependence of the phonon spectrum from where we extract relevant information
about the spin-gap ground state below $T_{c1}$.

\section{Experiment and Results}
Our $TiOCl$ single crystals were synthesized by standard vapor-transport
techniques from $TiO_{2}$ and $TiCl_{3}$, according to the procedure
described in Ref. \onlinecite{schaefer}. The crystals were checked by X-ray
diffraction experiments and static magnetic susceptibility
\cite{seidel,imai}. $TiOCl$ is an oxyhalogenide with layered structure formed
of $Ti^{3+}O^{2-}$ bi-layers, separated by $Cl^{-}$ bi-layers. The basic
$TiO_{4}Cl_{2}$ octahedra build an edge-shared network in the ab-plane of
the orthorhombic unit cell. $\chi(T)$, besides the already mentioned kink and sharp drop at
$T_{c2}$ and $T_{c1}$ respectively, is also characterized by a broad maximum
around 400 $K$. Above 100 $K$ $\chi(T)$ can be fitted by a S=1/2 Heisenberg
spin chain model with an antiferromagnetic (AF) exchange coupling constant $J=660$
$K$ (Ref. \onlinecite{seidel}). Furthermore, the specific heat $C_{p}(T)$
displays an anomaly only at $T_{c2}$ (Ref. \onlinecite{lee}).

We have measured the optical reflectivity $R(\omega)$ from the
far infrared (FIR) up to the ultra-violet (UV) spectral range at
temperatures between 10 and 300 $K$ and also as a function of magnetic 
field. Since we did not find any magnetic field dependence, we will focus 
on the temperature dependence only. The Kramers-Kronig (KK)
transformation of $R(\omega)$ allows us to evaluate the optical functions, like the real
part $\sigma_{1}(\omega)$ of the optical conductivity. Further details
pertaining to the experimental method can be found elsewhere \cite{wooten}. Light was
linearly polarized along the chain b-axis and
the transverse a-axis \cite{seidel,beynon}. In order to avoid leakage
effects of the polarizer, the polarization of light in our experiment always
coincides with the vertical axis of the sample mounting so that the
investigated crystallographic direction was perfectly parallel to the polarization
of the light beam. Therefore, the polarization dependence was
obtained by rotating the sample (instead of the polarizer) by 90$^{0}$ degrees
inside the cryostat. This assures that no undesired projections of
the light polarization along any transverse crystallographic
direction occurs in our experiment.

\begin{figure}[t]
   \begin{center}
    \leavevmode
    \epsfxsize=1\columnwidth \epsfbox {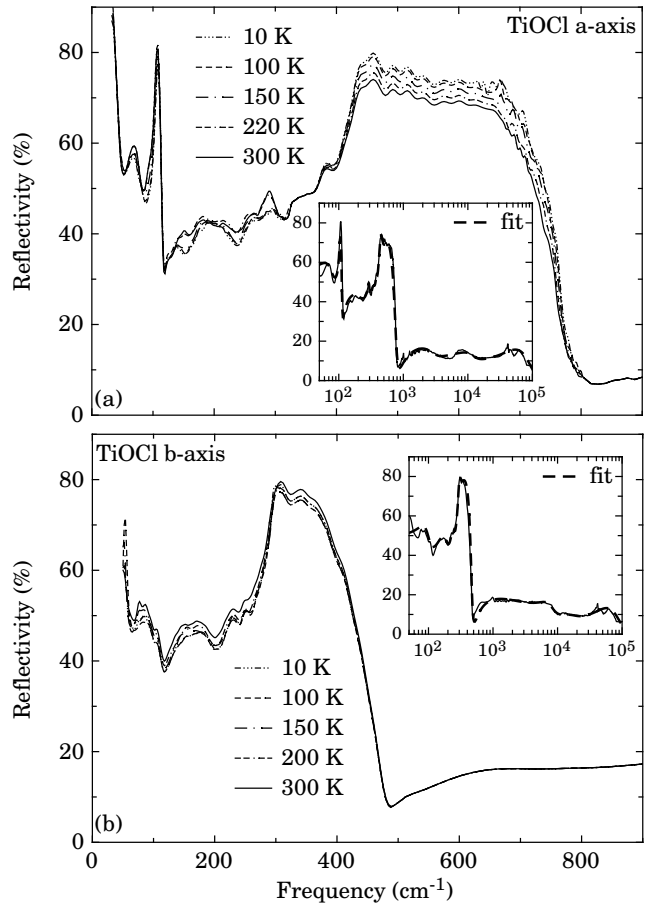}
     \caption{Optical reflectivity $R(\omega)$ in the infrared
spectral range of $TiOCl$ along the
     a-axis (a) and b-axis (b). The insets show the whole spectra at
300
     $K$ up to the ultraviolet spectral range and the fit as described
in the
text.}
\label{refl}
\end{center}
\end{figure}

Figure 1 summarizes our results, by focusing the attention on the
temperature dependence of $R(\omega)$ in the infrared spectral range
and for both polarization directions. The insets display the whole
$R(\omega)$ spectra at 300 $K$ with a logarithmic energy scale.
The first obvious observation is the strong anisotropy of the optical
response within the ab-plane and for photon energies below $\sim 10^{4} 
~cm^{-1}$. Despite the two dimensional layered-like 
structure, the anisotropy in the lattice dynamics may be 
considered as a signature for the low dimensionality of
$TiOCl$. Figure 2 reproduces the real part $\sigma_{1}(\omega)$ of
the optical conductivity at 300 $K$ in the far infrared spectral
range, as obtained from the KK
transformation of $R(\omega)$. The inset in panel (b) enhances the
visible spectral range, stressing the feature around 8x10$^{3}$
$cm^{-1}$ ($\sim$ 1 $eV$) for both polarization directions. Above 4x10$^{4}$ $cm^{-1}$
the $\sigma_{1}(\omega)$ spectra (not shown here, see Ref.
\onlinecite{lemmens} for more details) were found to be polarization
independent.

\begin{figure}[t]
   \begin{center}
    \leavevmode
    \epsfxsize=1\columnwidth \epsfbox {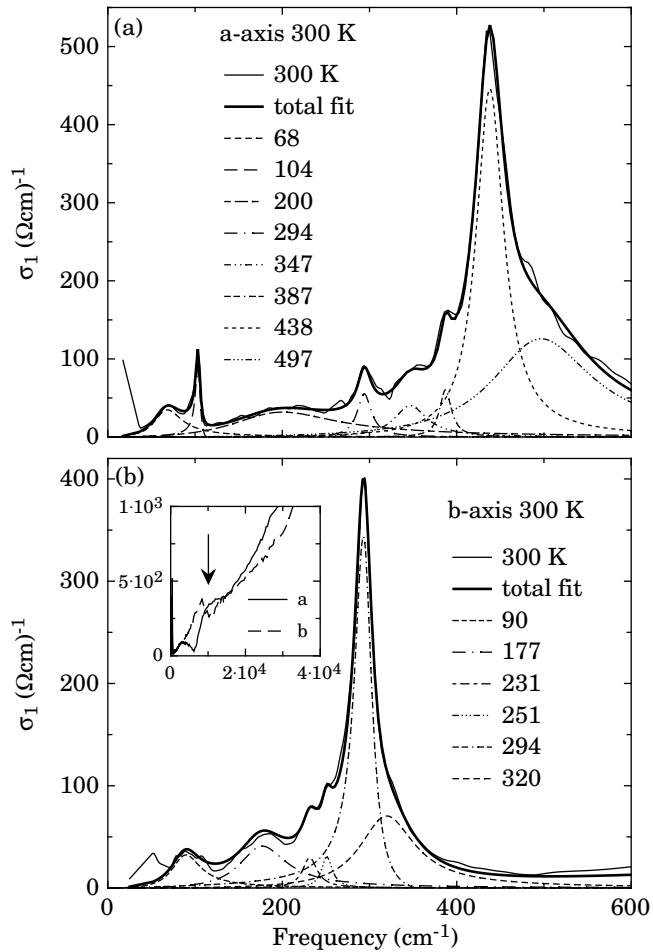}
     \caption{Real part $\sigma_{1}(\omega)$ of the optical
     conductivity at 300 $K$ in the infrared spectral range of $TiOCl$
along the a-axis (a) and b-axis (b). The total fit and its
components,
identified in the legend by their respective resonance frequency in
$cm^{-1}$, are also
     shown \cite{param}. The
     inset enlarges the visible spectral range and the arrow indicates
     the feature at $8100 ~cm^{-1}$ (1 $eV$) for both polarizations (see text). The
     temperature dependence of $\sigma_{1}(\omega)$ was already shown
     in Ref. \onlinecite{lemmens}.}
\label{sigma1}
\end{center}
\end{figure}

\section{Discussion}
The high frequency part of the excitation spectrum \cite{lemmens} is
dominated by electronic interband transitions (Fig. 1 and inset Fig.
2b). Recent LDA+U calculations \cite{seidel}, using the full-potential LMTO
method, predict a split-off of the (one-dimensional) t$_{2g}$ bands
creating an insulating state with a (charge) gap of about $8100 ~cm^{-1}$ (1
$eV$). The t$_{2g}$ band is derived from the d$_{xy}$ orbitals 
corresponding to the linear $Ti$-chains along the crystallographic b-axis 
\cite{seidel}. This agrees with our data displaying a peak at $\sim 8$x10$^{3} ~cm^{-1}$
along the chain b-axis and a pronounced shoulder at the same energy along
the transverse a-axis (see arrow in inset of Fig. 2b). The same band
structure calculations \cite{seidel} suggest furthermore interband
transitions between the $O$ and $Cl$ p-levels and the $Ti$ d-levels at
energies between 3.2x10$^{4}$ and 5.6x10$^{4} ~cm^{-1}$ (4 and 7 $eV$). This
is again very much in agreement with the absorption features seen in our
$\sigma_{1}(\omega)$ spectra (see insets of Fig. 1 in Ref.
\onlinecite{lemmens}) along both polarization directions.

\begin{figure}[t]
   \begin{center}
    \leavevmode
    \epsfxsize=1\columnwidth \epsfbox {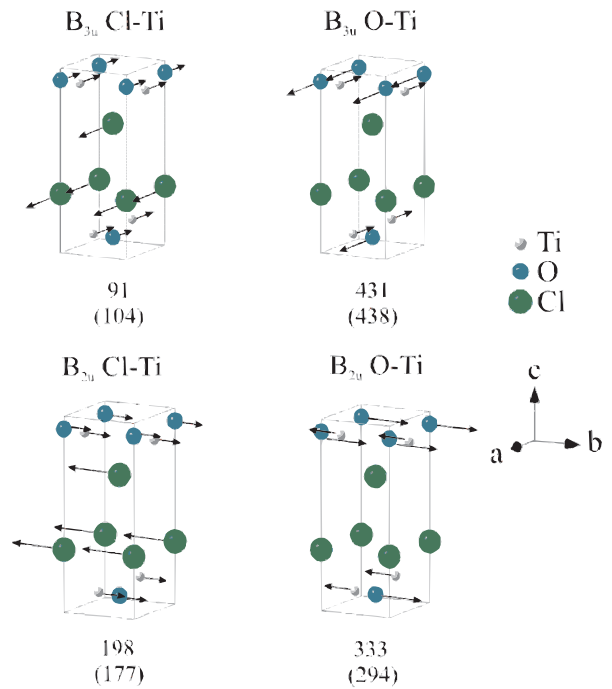}
     \caption{Schematic representation of the eigenvectors for the $B_{3u}$ and
$B_{2u}$ normal modes in $TiOCl$. The atom displacements for the IR active phonons occur within the
     ab-plane. The calculated
normal frequencies in $cm^{-1}$ are compared with the observed values (in
brackets).}
\label{phonon}
\end{center}
\end{figure}

In the far infrared spectral range (main panels of Fig. 1), several
absorptions dominate the $R(\omega)$ spectra. The absorption features are
better seen in the real part $\sigma_{1}(\omega)$ of the optical
conductivity (Fig. 2). On the one hand, along the chain b-axis (Fig. 2b)
there is a strong peak at 294 $cm^{-1}$ with additional absorptions,
overlapped to its low frequency tail, at 251 and 231 $cm^{-1}$. Moreover, we
recognize broad absorptions at 177 $cm^{-1}$ and around 90 $cm^{-1}$. Along
the transverse a-axis (Fig. 2a) on the other hand, there is a strong peak at
438 $cm^{-1}$ and less intensive absorptions at 68, 104, 294, 347 and 387
$cm^{-1}$, as well as a very broad feature around 200 $cm^{-1}$. A few
absorptions are furthermore characterized by an asymmetric shape (see
below). The strongest modes at 294 and 438 $cm^{-1}$ along the b- and
a-axis, respectively, display rather broad high frequency tails, which might
be indicative of some anharmonicity.

For $TiOCl$ with the bisandwich layer structure of $FeOCl$-type
\cite{beynon}, the space group $Pmmn (59, D_{2h})$ can be considered at room
temperature, for which two $B_{3u}$ modes polarized along the a-axis, two
$B_{2u}$ along the b-axis and two $B_{1u}$ along the c-axis can be predicted
as infrared (IR) active phonons. Because our samples are rather thin, we
only have access to the $B_{2u}$ and $B_{3u}$ modes of the ab-plane. We note that 3$A_{g}$, 3$B_{2g}$ and 3$B_{3g}$ Raman active phonons
are expected, as well. For light polarized within the ab-plane only the
3$A_{g}$ modes, inducing displacements along the c-axis, can be detected
\cite{lemmens}. Classical shell model calculations allow to extract
eigenfrequencies and eigenvectors for both Raman and IR-active phonons of
$TiOCl$ (Ref. \onlinecite{kovaleva}). As far as the IR phonons are 
concerned, the calculations predict the $B_{3u}$
(a-axis) phonons at 91 and 431 $cm^{-1}$ and the $B_{2u}$ (b-axis) at 198
and 333 $cm^{-1}$. The agreement between the calculated phonon frequencies
and the experimental observation is pretty good. Particularly, the two high frequency ones can be identified
with the most pronounced features in the experimental a- and b-axis spectra, 
respectively. Figure 3 shows the normal mode eigenvector patters for the IR 
active phonons, resulting from our shell model calculations.
Table I summarizes the predicted and the so far experimentally determined 
phonon frequencies for the IR-active modes and, as complement, for the 
Raman-active modes \cite{lemmens}, as well.

\begin{table}
\caption{\label{tab:phonon} Infrared and Raman active phonon for 
$TiOCl$ after the space group $P_{mmn}(59,D_{2h})$. The calculated 
(shell model \cite{kovaleva}) mode frequencies in $cm^{-1}$ are compared to the 
experimentally obtained values (in brakets). $A_{g}$ modes are observable in
(aa) or (bb) polarization. $B_{2g}$ and $B_{3g}$ modes are only accessible
in (ac) and (bc) polarization, respectively, with one polarization vector
parallel to the c axis.}
\begin{ruledtabular}
 \begin{tabular}{l l l l l l }
  \multicolumn{3}{c}{\bf{Infrared}} & \multicolumn{3}{c}{\bf{Raman}}\\
  \bf $B_{1u}$ & \bf $B_{2u}$ [a] & \bf $B_{3u}$ [a] & \bf $A_{g}$ [b] & \bf $B_{2g}$ & \bf $B_{3g}$\\
\hline
  308 & 198 & 91 & 248 & 84 & 126\\
  - & (177) & (104) & (203) & - & -\\
  433 & 333 & 431 & 333 & 219 & 237\\
  - & (294) & (438) & (365) & - & -\\
  - & - & - & 431 & 491 & 390\\
  - & - & - & (430) & - & -\\
\end{tabular}
\bf[a] this work\\
\bf[b] Ref. \onlinecite{lemmens}\\
\end{ruledtabular}
\end{table}

Even though the calculated frequencies of the IR phonon modes agree well with some features in
our spectra, a larger number of phonon-like absorptions than theoretically
predicted is found for both polarizations. The sample mounting chosen in our
experiment (see above) allows us to exclude leakage effects due to the
polarizer. However, one might invoke some twinning of the specimens. This
could lead to additional phonon modes, due to the mixing of different
polarization directions. This possibility is, nevertheless, rather unlikely
because of the strong anisotropy of the optical spectra. If domains would be
present because of twinning, phonons for both crystallographic axes would be
detected for both polarizations of light and no anisotropy would be found.
For instance the peaks at 387 and 438 $cm^{-1}$ along the a-axis (Fig. 2a)
are totally absent in the spectra along the b-axis (Fig. 2b). The same
applies for the sharp feature at about 104 $cm^{-1}$ for the a-axis.
Moreover as it will be discussed in details below, the temperature
dependence in $\sigma_{1}(\omega)$ can differ substantially even for
absorptions coinciding  at the same frequency in both polarizations.
Therefore, we are confident that no common features are shared in FIR among
the two polarizations. We propose that the effective space group
corresponds to a lower symmetry than assumed so far, or that additional
surface modes may become IR-active.

In order to shed light on the temperature dependence of the phonon
spectrum we have fitted the optical conductivity with the so-called
Fano expression \cite{cardona,fano}:
\begin{equation}
\tilde{\sigma}(\omega)=\sum_{j}i\sigma_{0j}\frac{(q_{j}+i)^{2}}{i+x(\omega)},
\end{equation}
with $x(\omega)=\frac{\omega_{0j}^{2}-\omega^{2}}{\gamma_{j}\omega}$, where
$\omega_{0j}$ is the resonance frequency, $\gamma_{j}$ is the width (i.e.,
damping) and $\sigma_{0j}=\omega_{pj}^{2}/\gamma_{j} q_{j}^{2}$ with
$\omega_{pj}$ as the oscillator strength and $q_{j}$ as the so-called
asymmetry factor of the $j$-absorption. The asymmetric line shape for the
(sharp) phonon modes derives from an interaction between lattice vibrations
and a continuum, usually given by an electronic background. The 
interaction with a magnetic continuum may also lead to a Fano lineshape, as 
shown in the Raman scattering spectra of low dimensional spin systems 
\cite{lemmensrev,choi}. It is
verified easily that for the dimension-less Fano parameter $q_{j}\to\infty$ one can
recover the lineshape of the harmonic Lorentzian oscillator
\cite{wooten,cardona}. The approach of eq. (1) was successfully applied by
Damascelli {\it et~al.} \cite{damascelli2,damascelli3} to the spin-Peierls system
$\alpha^{'}-NaV_{2}O_{5}$.

Alternative approaches, based on the Fano theory \cite{fano}, are available
in the literatures. Here, we quote first of all the work by Lupi {\it
et~al.} \cite{lupi} on the phonon interaction with a polaronic background in
$Nd_{1.96}Ce_{0.04}CuO_{4+y}$. This approach is based on the extension of
Fano's formalism by Davis and Feldkamp (DF) \cite{davis}, who considered the
case of an interaction between an arbitrary number of discrete states and
continua. In this work, we will primarily focus our attention on the fits of
$\sigma_{1}(\omega)$ after eq. (1). Nevertheless, in order to stress the
equivalence among different approaches, we will discuss and compare the
relevant asymmetry $q_{j}$-factors obtained with eq. (1) and by the DF
formalism. Secondly, we also mention the discussion of the c-axis phonon
modes in $YBa_{2}Cu_{3}O_{y}$ by Sch\"utzmann {\it et~al.}
\cite{schuetzmann}. In this phenomenological approach, the conventional
Lorentz shape was modified by a factor $e^{i\theta}$, where $\theta$
accounts for the asymmetry. This model, however, does not add relevant
physical insight to the discussion with respect to the more simple Lorentz
model \cite{wooten}. Therefore, it will not be considered further.

The total fit of $\sigma_{1}(\omega)=Re(\tilde{\sigma}(\omega))$, covering
the whole spectral range from FIR up to UV, is obtained by summing over
eleven and ten contributions in eq. (1) for the a- and b-axis (Ref.
\onlinecite{param}), respectively. The fit of $\sigma_{1}(\omega)$ at 300
$K$ and the single components in FIR for both polarization directions are
shown in Fig. 2. The reproduction of the experimental $\sigma_{1}(\omega)$
curve is astonishingly good and the same fit quality is obtained at all
temperatures. In passing, we also note that the same set of fit parameters
\cite{param} allows us to reproduce the measured $R(\omega)$ spectra (see,
e.g., the fit of $R(\omega)$ at 300 $K$ in the insets of Fig. 1).

\begin{figure}[t]
   \begin{center}
    \leavevmode
    \epsfxsize=.8\columnwidth \epsfbox {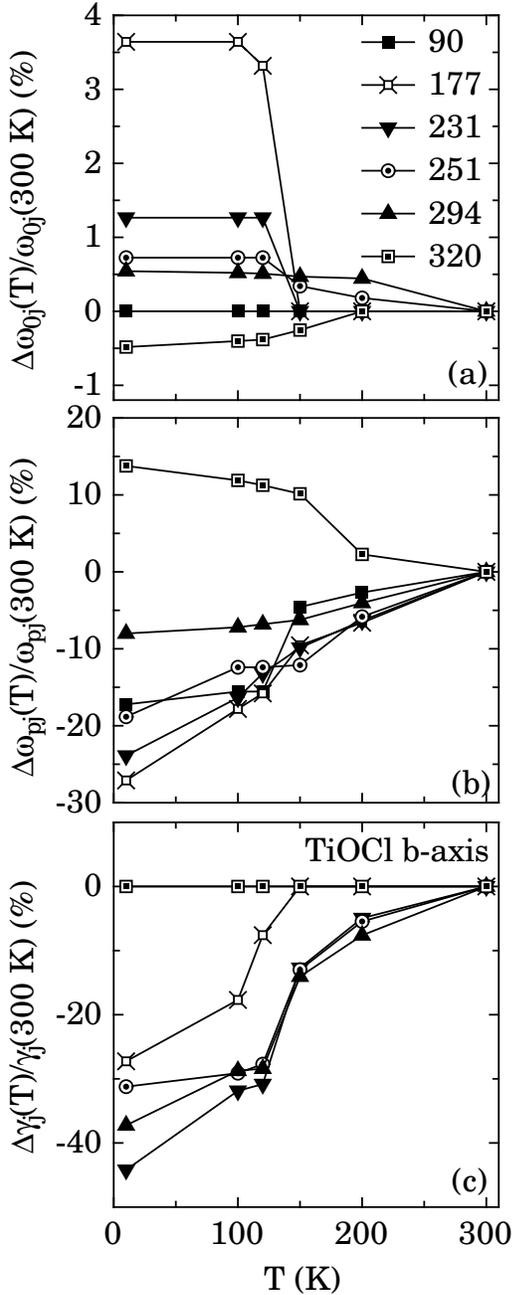}
     \caption{Temperature dependence along the b-axis of the percentage change with
respect to 300 $K$ (see text) for the resonance frequencies
($\omega_{0j}$) (a), the
     oscillator strengths ($\omega_{pj}$) (b) and the dampings
($\gamma_{j}$) (c)
of the phonon modes (identified in the legend by their respective
resonance frequency in $cm^{-1}$).}
\label{paramb}
\end{center}
\end{figure}

\begin{figure}[t]
   \begin{center}
    \leavevmode
    \epsfxsize=.8\columnwidth \epsfbox {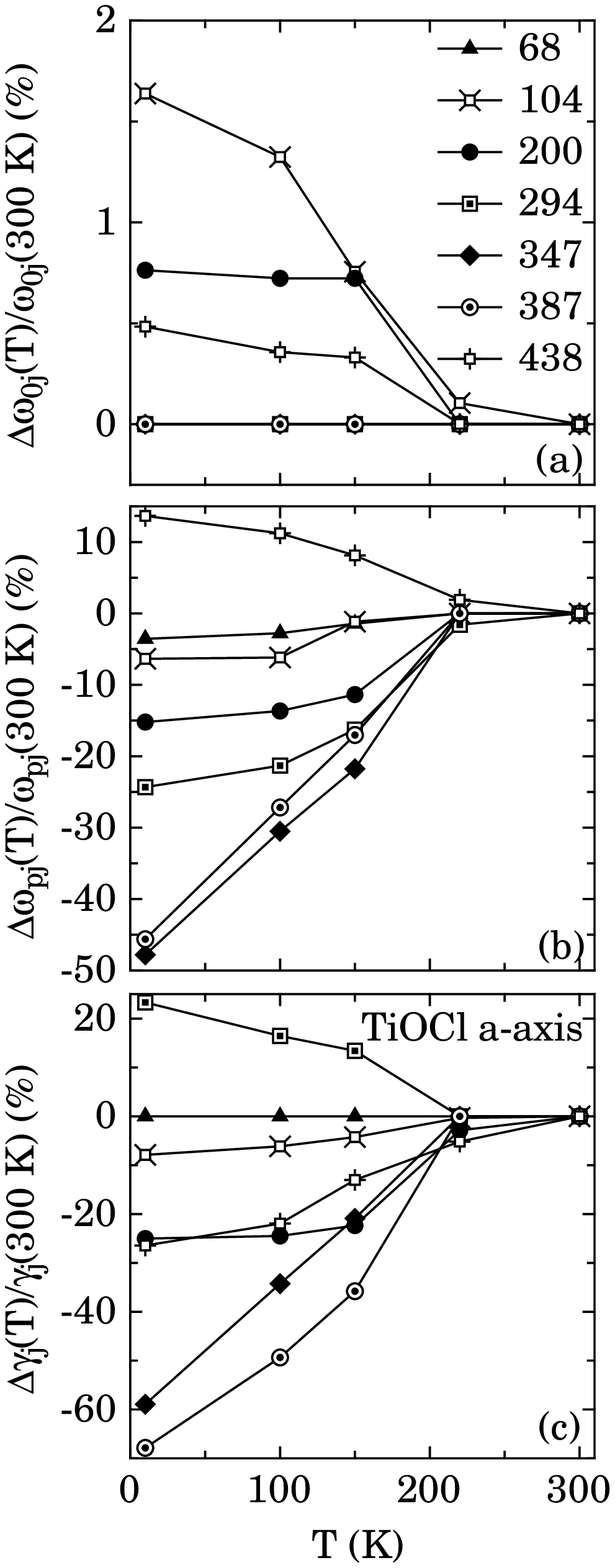}
     \caption{Temperature dependence along the a-axis of the percentage change with
respect to 300 $K$
     (see text) for the resonance frequencies ($\omega_{0j}$) (a), the
     oscillator strengths ($\omega_{pj}$) (b) and the dampings
($\gamma_{j}$) (c)
of the phonon modes (identified in the legend by their respective
resonance frequency in $cm^{-1}$).}
\label{parama}
\end{center}
\end{figure}

The temperature dependence of the fit parameters ($\omega_{0j}, \gamma_{j},
\omega_{pj}$ and $q_{j}$) \cite{param} is shown, as percentage change with
respect to the 300 $K$ data (e.g., $\Delta\omega_{0j}(T)/\omega_{0j}(300
~K)$, with $\Delta\omega_{0j}(T)=\omega_{0j}(T)-\omega_{0j}(300 ~K)$), in
Figs. 4, 5 and 6 for both polarizations. Along the a-axis the lower seven
while along the b-axis the lower six oscillators are temperature dependent
and will be here discussed further. The overall temperature dependence
mainly develops below 200 $K$ and tends to saturate below 100 $K$ 
(particularly for the chain b-axis). The
temperature dependence, occurring in a broad temperature interval and
extending well above $T_{c1}$, is indicative of an extended fluctuation
regime, which has been recognized in NMR data \cite{imai}, as well. The
relaxation rates of $^{35}Cl$ sites show dynamic lattice distortion with
onset at 200 $K$ while for the $^{46,49}Ti$ sites, $1/TT_{1}$, which probes
the spin degrees of freedom, forms a maximum at about 135 $K$ (Ref.
\onlinecite{imai}). Therefore, the interplay between the lattice and
spin degrees of freedom must be already taking place at high
temperatures. The temperature dependence of $1/TT_{1}$ implies
a pseudo-gap phase in the homogeneous state of the spin system with an
estimated pseudo-gap $\Delta_{fluct}\sim 430 ~K$ (Ref. \onlinecite{imai}).
The temperature dependence of the ESR parameters (i.e., the ESR line width
$\Delta H$ and the so-called ESR $g$-tensor) also displays a progressive
evolution over a broad temperature range between 200 $K$ and $T_{c1}$ (Ref.
\onlinecite{kataev}). From the analysis of the ESR signal it is claimed that
a strong coupling between spin and lattice degrees of freedom exists and
that spin as well as orbital fluctuation effects above $T_{c1}$ may be
responsible for the peculiar temperature dependence of various quantities
(e.g., $\chi(T)$).

The resonance frequencies ($\omega_{0j}$) of almost all phonons tend to
increase (Fig. 4a and 5a), though moderately (i.e., the change does not
exceed 4 $\%$), with decreasing temperature, indicating a progressive
hardening of the modes. The resonance at 320 $cm^{-1}$ along the b-axis,
accounting for the broad high frequency tail of the mode at 294 $cm^{-1}$,
displays on the contrary a weak softening. A well-defined soft mode is in
principal expected in models for the conventional SP transition where the
structural deformation is driven by a linear coupling between the lattice
and the magnetic degrees of freedom. As far as the dynamical interplay
between spins and phonons in $TiOCl$ is concerned, it is clear from the
temperature dependence of our optical spectra that neither a soft mode nor a
generalized red-shift in the resonance frequency of the phonons have been
detected. At this point, apart from the 320 $cm^{-1}$ mode along the b-axis,
we speculate that the transition at $T_{c1}$ is not driven by a softening of
the IR phonon spectrum. A final word
on the soft mode issue will only be possible with neutron scattering
experiments, since the lattice dimerization involved in the Peierls
transition is related to normal modes away from the Brillouin zone center.
From the perspective of the IR spectrum, one could have hoped that the
presence of a soft mode would have resulted in an overall softening of the
phonon-branch it belongs to. The absence of a clear signature for IR phonons
softening is common to other spin-Peierls systems, like $CuGeO_{3}$ (Ref.
\onlinecite{damascelli}). Nevertheless, evidence for phonon softening has 
been found in the Raman scattering experiments \cite{lemmens}. A broad 
feature was identified at about 160 $cm^{-1}$ for the chain b-axis 
polarization. The weight of this signal
gets confined and softens down to 130 $cm^{-1}$, i.e, by $\sim 20\%$, with
decreasing temperature. This mode is ascribed to a Brillouin zone boundary
phonon. The experimentally observed $A_{g}$ mode at 203 $cm^{-1}$ (Table I) 
with light polarization along the b-axis is
its related $\Gamma$-point Raman-allowed phonon \cite{lemmens}.

Most of the phonon modes get narrow
with decreasing temperature (Fig. 4c and 5c). Only the mode at 294 $cm^{-1}$ along the a-axis displays a broadening with 
decreasing temperature. Particularly the width
$\gamma_{j}$ of quite all phonons along the chain b-axis follows very closely the temperature
dependence of the magnetic susceptibility $\chi(T)$. The temperature 
dependence of $\gamma_{j}$ in $TiOCl$ is very similar to the findings in 
$\alpha^{'}-NaV_{2}O_{5}$ (Refs. \onlinecite{damascelli2}). The pronounced 
narrowing of the modes occur in the temperature interval of the so-called 
pseudo spin-gap phase identified in the NMR spectra \cite{imai}. It seems 
therefore natural to relate this phonon narrowing with the suppression of 
low frequency spin fluctuations and to consider it as another fingerprint 
for the coupled spin-lattice fluctuations.

\begin{figure}[t]
   \begin{center}
    \leavevmode
    \epsfxsize=1\columnwidth \epsfbox {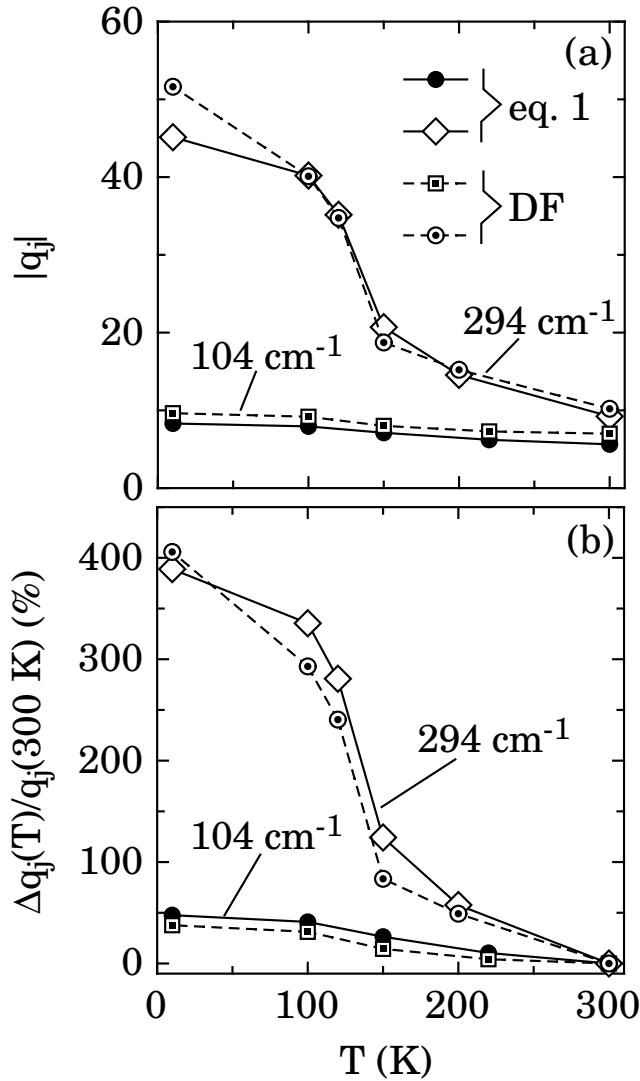}
     \caption{(a) Temperature dependence of the asymmetry
     factor $\mid q_{j}\mid$ for the 104 $cm^{-1}$ mode along the
a-axis
     and the 294 $cm^{-1}$ mode along the b-axis,
     calculated after eq. (1). Note that $q_{j}<0$ for both
     polarization directions. (b) Temperature dependence of the
corresponding percentage changes with
respect to 300 $K$ (i.e., $\Delta q_{j}(T)/q_{j}$(300
     $K$), with $\Delta q_{j}(T)=q_{j}(T)-q_{j}$(300 $K$)) for the
asymmetry factor $q_{j}$ of
     the a- and b-axis. In addition, we also display the
     $q_{j}$-factor and its percentage change for both modes obtained
with the Fano
     formalism, based on the approach of Davis and Feldkamp (DF)
\cite{davis}.
     The equivalence of the two approaches (eq. (1) and DF) is
obvious.}
\label{paramq}
\end{center}
\end{figure}

It turns out that most of the absorptions, seen in our spectra (Fig. 1 and
2) and described by the $j$-components in eq. (1), reduce to the
Lorentzian-like (i.e., $q_{j}\to\infty$) shape. Only the peak at 104
$cm^{-1}$ along the a-axis and at 294 $cm^{-1}$ along the b-axis display a
Fano-like asymmetry. The asymmetry (Fig. 6) of the mode at 104 $cm^{-1}$
along the a-axis gradually decreases (i.e., $\mid q_{j}\mid$ gets larger)
with decreasing temperature, although the mode remains considerably
asymmetric at all temperatures. This indicates that there is a predominant
interaction with the continuum both above and below $T_{c1}$. On the
other hand, the temperature dependence of the asymmetry for the 294
$cm^{-1}$ mode along the b-axis displays a clear crossover between
200 $K$ and $T_{c2}$ from an asymmetric Fano-like shape (i.e., $\mid
q_{j}\mid$ small) to a Lorentzian oscillator (i.e., $\mid q_{j}\mid$ very
large). The distinct behaviour in the temperature dependence of the 
$q_j$-factors within the ab-plane is a further fingerprint for the 
anisotropy of the lattice dynamics as well as of the coupling between 
phonon and continuum. The clear Fano-Lorentz cross-over
along the chain b-axis suggests moreover the suppression of the interaction
with the continuum with decreasing temperature \cite{damascelli2}. This, combined with the 
anisotropic temperature dependence of the $q_j$-factors themselves, hints 
once more to a progressive dimensionality cross-over (two-to-one) within 
the ab-plane with decreasing temperature. Since $q_{j}<0$ for both asymmetric 
modes, the relevant continuum of excitations, that is renormalized with 
temperature, covers an energy interval below
the phonon resonance frequencies. For the chain b-axis, this identifies a 
characteristic energy scale of the order of 400 $K$. In Fig. 6, we also report,
for comparison, the $q_{j}$-factor for both modes as calculated from the
Fano model based on the approach of Davis and Feldkamp \cite{davis}. Even
though the two Fano approaches \cite{cardona,davis} (eq. (1) and DF) are
formally different (i.e., they are characterized by different energy
powerlaw decays of the absorption coefficient), the corresponding
$q_{j}$-factors are identical both in absolute value \cite{param} and in the
relative percentage change (Fig. 6). This stresses the equivalence of the
Fano asymmetry concept (parameterized by the $q_{j}$-factors) for both
calculations.

The temperature dependence of the spectral weight encountered in the phonon
spectrum is also of interest. We have observed \cite{lemmens} that the
spectral weight encountered in $\sigma_{1}(\omega)$ at low frequencies tends
to decrease below 200 $K$. The total spectral weight, obtained by
integrating $\sigma_{1}(\omega)$ up to the UV spectral range, however, is
conserved and it is fully recovered already by $10^{4} ~cm^{-1}$ ($\sim$ 1
$eV$) \cite{lemmens}. This is confirmed by the behaviour of the squared ($\omega_{pj}^{2}$)
oscillator strengths (Fig. 4b and 5b). Particularly for the b-axis, 
the spectral weight is progressively removed with decreasing temperature 
over a spectral range of roughly 400 $K$. This is 
in agreement with the energy scale identified above in the analysis of the 
phonon asymmetry within the Fano approach.
Most of the suppressed weight at low frequencies moves into the modes at 294
or 438 $cm^{-1}$ or into their high frequency tails along the b- and a-axis,
respectively. The remaining part of the suppressed weight shifts from the
elastic degrees of freedom towards zone boundary (folded) modes or, as we
favor for $TiOCl$, to the electronic degrees of freedom at high energies.
Again this overall shift mainly occurs between $T_{c2}$ and 200 $K$.

Finally, the giant reduction of the phonon width, the behaviour of the Fano
$q_{j}$-factors (particularly along the b-axis) as well as the suppression
of spectral weight in FIR (Figs. 4, 5 and 6) suggest the presence and
development of a characteristic energy scale. Above
all the depletion of spectral weight over an energy range of the order of 400 $K$ is very much reminiscent of a similar
behaviour in the Raman spectra, occurring over the same energy interval with
decreasing temperature and associated to the spin-gap opening
\cite{lemmens}. Setting $2\Delta_{opt}\sim$ 300 $cm^{-1}$ ($\sim 430$ K) as
lower bound for the spin-gap energy, we obtain a reduced gap ratio
$2\Delta_{opt}/k_{B}T_{c1}\sim 4.6$ and 6.7 for $T_{c2}$ and $T_{c1}$,
respectively. With respect to the mean-field results (i.e., 
$2\Delta/k_{B}T_{c}\sim$ 3.52) our larger gap ratios might reveal 
competing exchange paths or electronic degrees of freedom \cite{lemmens}, 
but are more reasonable than those obtained from recent NMR study 
\cite{imai}. Indeed, the extraordinary larger
spin-gap $\Delta_{fluct}$ would correspond to a reduced gap
ratio ranging between 9 and 13 for $T_{c1}$ and $T_{c2}$, respectively. One
can reconcile the outcome of a typical local probe like NMR with the optical
results, which notably give an average perspective of the excitation
spectrum, by following the development of both the magnetic and structural
correlations. With decreasing temperature down to $T_{c2}$ the coherence
length of the structural distortion increases and the magnetic correlations
cross-over from 2D to 1D (Ref. \onlinecite{lemmens}). The energy gain at
$T<T_{c2}$ is mainly related to the spin system. Therefore, the related
anomaly in the specific heat \cite{lee} is small and in conventional X-ray
scattering no sign of a coherent structural distortion can be found
\cite{kataev}. With further decreasing $T$ the structural distortion becomes
long range. Below $T_{c1}$, $TiOCl$ has a conventional behaviour closely
related to a spin-Peierls system. Therefore, the spin-gap $2\Delta_{fluct}$
(from NMR \cite{imai}) is the lowest energy for the local double spin-flip
in the short range order distorted phase, while $2\Delta_{opt}$ (from IR and
Raman \cite{lemmens}) is the energy for the global spin-gap of the fully
dimerized chain.

\section{Conclusion}
The phase transition of $TiOCl$ at $T_{c1}$ is not of simple character and
several features point towards an unconventional spin-gap formation.
Particularly intriguing is the temperature dependence of the phonon spectrum
which, alike the NMR and ESR spectra, and the magnetic
susceptibility, hints towards the key role of both orbital and magnetic
degrees of freedom in shaping the magnetic properties of $TiOCl$.
Fluctuation effects, extending over a broad temperature range from $T_{c1}$
up to 200 $K$, seem to be the most peculiar property of this low-dimensional
spin-1/2 quantum system and may be indicative of a novel coupling between
low and high energy scales.

\acknowledgments
The authors wish to thank J. M\"uller for technical help, and A.
Damascelli, B. Schlein, P. Calvani and A.
Perucchi for fruitful discussions. This work
has been supported by the Swiss National Foundation for the
Scientific Research, INTAS 01-278, DFG SPP1073 and by the MRSEC Program of
the National Science Foundation under award number DMR 02-13282.

\newpage

\end{document}